\documentclass[
superscriptaddress,secnumarabic,amssymb,amsmath,
nobibnotes,aps,prd,nofootinbib,nopacsnumber,onecolumn,10pt
]{revtex4}
\usepackage{graphicx}
\usepackage{bm}
\usepackage{amsmath}
\usepackage{amsfonts}
\usepackage{amssymb}
\usepackage[utf8]{inputenc}%
\providecommand{\U}[1]{\protect\rule{.1in}{.1in}}

\newcommand{\be}{\begin{equation}}
\newcommand{\ee}{\end{equation}}

\newcommand{\mincir}{\raise
-3.truept\hbox{\rlap{\hbox{$\sim$}}\raise4.truept\hbox{$<$}\ }}
\newcommand{\magcir}{\raise
-3.truept\hbox{\rlap{\hbox{$\sim$}}\raise4.truept\hbox{$>$}\ }}

\begin{document}
\title{Bianchi I spacetimes within 4D Einstein-Gauss-Bonnet scalar field theory}
\author{Alex Giacomini}
\email{alexgiacomini@uach.cl}
\affiliation{Instituto de Ciencias Fisicas y Matem\`{a}ticas, Universidad Austral de Chile,
Valdivia, Chile}
\author{Chevara Hansraj}
\email{hansrajc@sun.ac.za}
\affiliation{Applied Mathematics Division, Department of Mathematical Sciences,
Stellenbosch University, Private Bag X1, Matieland 7602, South Africa}
\affiliation{National Institute for Theoretical and Computational Sciences (NITheCS),
Stellenbosch, South Africa}
\author{Genly Leon}
\email{genly.leon@ucn.cl}
\affiliation{Departamento de Matem\`{a}ticas, Universidad Cat\`{o}lica del Norte, Avda.
Angamos 0610, Casilla 1280 Antofagasta, Chile}
\affiliation{Institute of Systems Science, Durban University of Technology, Durban 4000,
South Africa}
\author{Andronikos Paliathanasis}
\email{anpaliat@phys.uoa.gr}
\affiliation{Institute of Systems Science, Durban University of Technology, Durban 4000,
South Africa}
\affiliation{Centre for Space Research, North-West University, Potchefstroom 2520, South Africa}
\affiliation{Departamento de Matem\`{a}ticas, Universidad Cat\`{o}lica del Norte, Avda.
Angamos 0610, Casilla 1280 Antofagasta, Chile}
\affiliation{National Institute for Theoretical and Computational Sciences (NITheCS),
Stellenbosch, South Africa}

\begin{abstract}
We investigate the evolution of anisotropies in Bianchi I spacetimes within the framework of the 4D Einstein-Gauss-Bonnet scalar field theory. The field equations are formulated using dimensionless variables, and the asymptotic dynamics are studied through a combination of analytical and numerical techniques. For the locally rotationally symmetric case, we analytically explore the stationary points of the field equations. The analysis reveals the existence of accelerating solutions in which the scalar field and the Gauss-Bonnet scalar effectively play the role of a cosmological constant. As a result, both anisotropic and isotropic expanding solutions are recovered, along with the Minkowski spacetime. No scaling solutions are supported by the gravitational model. For the general anisotropic Bianchi I geometry with three distinct scale factors, we find that a class of compactified Kasner-like solutions is obtained. In addition, a new family of solutions follows, describing a two-dimensional splitting of the background geometry. This behavior is similar to the previously observed pure Einstein-Gauss-Bonnet theory in higher-dimensional spacetimes.  

\end{abstract}
\date{\today}
\maketitle

\section{Introduction}

The cosmological principle states that, on sufficiently large scales, the Universe is homogeneous and isotropic at the present epoch, in agreement with current observational data \cite{Hu:2001bc}. Nevertheless the study of the cosmic microwave background (CMB) suggests that the early Universe may have exhibited deviations from perfect isotropy \cite{Planck:2011ivn}. The mechanism responsible for the suppression of these anisotropies is cosmic inflation \cite{Guth:1980zm,Barrow:1993hn,Martin:2013tda,Achucarro:2022qrl,Barrow:2016wiy,DAgostino:2021vvv}. However, anisotropic effects are expected to have played an important role during the pre-inflationary epoch \cite{Barrow:1987ia}. 

Homogeneous and anisotropic four-dimensional spacetimes are classified into nine distinct geometric classes \cite{Ryan:1975jw}, corresponding to the Bianchi classification of three-dimensional Lie algebras. The simplest class is the Bianchi I geometry, in which the spacetime admits a three-dimensional Abelian Lie algebra generated by the translational symmetries that define the directions of the spatial hypersurface. 

In general relativity (GR), the gravitational field equations of the Bianchi I model reveal the Kasner universe \cite{Kasner:1921zz}, which is a closed-form vacuum anisotropic solution. The Kasner universe and its generalizations \cite{Kokarev:1995ri,Halpern:2002vd,Clifton:2005aj,Paliathanasis:2016vsw,Paliathanasis:2017htk,deCesare:2019suk,Delice:2004wk,Paliathanasis:2024qkh,Murtaza:2025gme,Toporensky:2016kss,Vitenti:2006hy,Leach:2006br} have various applications. It describes the chaotic behaviour of the mixmaster universe near the cosmic singularity \cite{Chernoff:1983zz}. Moreover, Kasner and Kasner-like geometries have been used as a paradigm for the study of the observational consequences of anisotropic expansion during inflation \cite{Soda:2012zm,Ghalee:2014goa,Nojiri:2022idp,Goodarzi:2022wli}, involving quantum particle creation \cite{Zeldovich:1971mw}, baryosynthesis \cite{Barrow:1981bv}, nucleosynthesis \cite{Rothman:1984vk} and many others \cite{Som:2002de,Halpern:2000xj,Krori:1996jv,Krori:1990gh,Caceres:2024edr,Arean:2024pzo,Brizuela:2024xyi,Bueno:2024fzg,Prihadi:2025czn,Fournodavlos:2020jvg}. 

Lovelock's theory of gravity \cite{Lovelock:1971yv} is an extension of general relativity in higher-dimensional geometries. It is a second-order theory free from Ostrogradsky instabilities
\cite{Woodard:2015zca} and recovers GR when the geometry reduces to four dimensions \cite{Padmanabhan:2013xyr}. In Lovelock theory, the gravitational field is described by a modified Einstein-Hilbert action, where new geometric scalars are introduced, which have the property of being topological invariants when the geometry has dimension four. The first nontrivial extension of the Einstein-Hilbert Action within Lovelock's theory is achieved with the introduction of the Gauss-Bonnet scalar \cite{Fernandes:2022zrq}. The Gauss-Bonnet term also appears naturally as the leading higher-curvature correction in low-energy effective actions motivated by string theory, where the ghost-free Gauss-Bonnet combination plays a central role \cite{Zwiebach:1985, Boulware:1985, Gross:1987, Kanti:2015}.

In four dimensions the Gauss-Bonnet scalar is a topological invariant and hence
does not contribute to the field equations dynamically. However the introduction of a nonminimally coupled scalar field $\phi$ to the Gauss-Bonnet scalar $G$ leads to the emergence of nontrivial dynamics in Einstein's gravity. The resulting theory is known as the Einstein-Gauss-Bonnet scalar field theory \cite{Tsujikawa:2006ph,Koivisto:2006xf,Carloni:2017ucm,Pozdeeva:2024ihc,Vernov:2021hxo,Pozdeeva:2021iwc,Millano:2023gkt,Millano:2024vju,Doneva:2019vuh,Nojiri:2023jtf,Paliathanasis:2024gwp}, where a particular case without a scalar field kinetic term is the $f\left( G\right)$-theory \cite{Nojiri:2005jg,Nojiri:2018ouv,Pozdeeva:2019agu,Li:2007jm,Nojiri:2021mxf,Nojiri:2024nlx,Dimakis:2025wtc,Elizalde:2019irv}. This framework leads to a wide variety of possible evolutionary paths and cosmic behaviors including inflationary and late-time accelerated epochs, as well as bouncing or singular evolutions \cite{Nojiri:2006, Tsujikawa:2007, Kanti:2015, Bamba:2015}.

Anisotropic spacetimes in $f\left(G\right)$-theory were investigated recently
in \cite{Bogadi:2025pur}. In particular, the cosmological dynamics of Bianchi I, Bianchi III, and the Kantowski-Sachs spacetimes were examined. It was found that,
regardless of the spatial curvature of the two-dimensional space, these
spacetimes admit as attractors a point that describes an isotropic and
spatially flat FLRW scaling solution, as well as the Minkowski spacetime, which
means that the theory can solve the flatness problem and can lead to an
isotropic universe. However, the model suffers from Big Rip and Big Crunch
singularities, which infers that the theory is affected by a fine-tuning
problem, as discussed previously in \cite{Papagiannopoulos:2025uix}. 

When a scalar field is coupled to the Gauss-Bonnet term, the new coupling changes the shear evolution. In many cases it can reduce the shear and drive the universe toward an isotropic (FLRW-like) state, but in other cases it can support sustained anisotropic phases \cite{Giacomini:2025hrd,Bogadi:2025pur,Do:2019,Naderi:2018}, which underlines the significance of studying these options in Bianchi I spacetimes to test how stable isotropic inflation really is. Nevertheless, in the Einstein-Gauss-Bonnet scalar field theory without a kinetic term \cite{Giacomini:2025hrd}, it was found that for a four-dimensional Bianchi I geometry, the resulting geometry leads to a splitting of the spatial part of the metric into a sum of isotropic two-dimensional and one-dimensional spaces. These solutions are unstable. Moreover, the isotropic universe is recovered, which is the attractor of the GR limit. 

In this work, in the same spirit, we investigate the evolution of anisotropies now in the presence of a kinetic term for the scalar field in the gravitational action integral. As we shall see, the kinetic term and the nature of the scalar field play important roles in the evolution of the anisotropies. In particular, while
the splitting of the geometry persists, this splitting can now become dynamically
preferred and even act as an attractor for broad classes of initial conditions, in
contrast to the purely non-kinetic case.

The structure of the paper is as follows. 
In Section \ref{sec2}, we briefly discuss the gravitational framework considered in this work, namely the four-dimensional Einstein-Gauss-Bonnet theory with a scalar field coupled to the Gauss-Bonnet topological invariant. We introduce the Bianchi I geometry as the simplest anisotropic cosmological model and present the corresponding minisuperspace formulation, together with the associated point-like Lagrangian.  In Section \ref{sec3}, we investigate analytically the asymptotic dynamics of the locally rotationally symmetric (LRS) Bianchi I spacetime. We find that the unique attractor solutions correspond to accelerated cosmic expansion in the absence of an explicit cosmological constant term. The most general anisotropic Bianchi I model is examined in Section \ref{sec4}, where the analysis is carried out using numerical methods. In addition to the accelerated solutions identified previously, we uncover a new class of Kasner-like solutions, which describe a $2+2$ compactification of spacetime. Finally, in Section \ref{sec5}, we summarize our results and compare them with those obtained in earlier studies.

\section{Einstein-Gauss-Bonnet Scalar Field Gravity}

\label{sec2}

In Einstein's GR, the gravitational field is described by the Ricci scalar of a four-dimensional Riemannian manifold. Consider the metric tensor $g_{\mu\nu}$ with covariant derivative $\nabla_{\mu}$ defined by the Levi-Civita connection $\Gamma_{\mu\nu}^{\kappa}$, i.e.%
\begin{equation}
\Gamma_{\mu\nu}^{\kappa}=\frac{1}{2}g^{\kappa\lambda}\left(  g_{\mu\lambda
,\nu}+g_{\lambda\nu,\mu}-g_{\mu\nu,\lambda}\right)  \text{,}%
\end{equation}
the curvature tensor, the Ricci tensor and the Ricci scalar are expressed in
terms of the connection as follows%

\begin{align}
R_{~\lambda\mu\nu}^{\kappa}  &  =2\partial_{\lbrack\mu}\Gamma_{\;\;\nu
]\lambda}^{\kappa}+2\Gamma_{\;\;[\mu|\sigma|}^{\kappa}\Gamma_{\;\;\nu]\kappa
}^{\sigma}~,\\
R_{\mu\nu}  &  =2\partial_{\lbrack\kappa}\Gamma_{\;\;\mu]\nu}^{\kappa}%
+2\Gamma_{\;\;[\kappa|\sigma|}^{\kappa}\Gamma_{\;\;\mu]\nu}^{\sigma}~,\\
R  &  =2g^{\mu\nu}\partial_{\lbrack\kappa}\Gamma_{\;\;\mu]\nu}^{\kappa
}+2g^{\mu\nu}\Gamma_{\;\;[\kappa|\sigma|}^{\kappa}\Gamma_{\;\;\mu]\nu}%
^{\sigma}~.
\end{align}

The introduction of the Gauss-Bonnet scalar \cite{Fernandes:2022zrq},%
\begin{equation}
G=R^{2}-4R_{\mu\nu}R^{\mu\nu}+R_{\mu\nu\sigma\lambda}R^{\mu\nu\sigma\lambda},
\end{equation}
in the Einstein-Hilbert action leads to the Einstein-Gauss-Bonnet theory of
gravity, that is \cite{Fernandes:2022zrq}%
\begin{equation}
S_{GB}=\int\sqrt{-g}\left(  R+\alpha_{0}G\right)  ,\label{sg1}%
\end{equation}
where $\alpha_{0}$ is a coupling constant. As mentioned previously, the latter scalar
does not contribute to the gravitational field equations when the geometry is
of dimensional four.

However, because the Gauss-Bonnet scalar is topologically invariant in the case
of a four-dimensional geometry, in \cite{Glavan:2019inb} the coupling constant $\alpha_{0}$ was proposed to depend on the dimension $n$ of the
background geometry, that is, $\alpha_{0}=\frac{\alpha_{1}}{n-4}$. Although 
such an approach can lead to pathologies as discussed in \cite{Arrechea:2020gjw}.

Two alternative approaches which have been considered in the literature to
introduce the effects of the Gauss-Bonnet scalar in the gravitational field
for a lower-dimensional geometry are introducing a nonlinear function $f\left(
G\right)$ in the action integral (\ref{sg1}) or considering the existence
of a scalar field nonminimally coupled to the Gauss-Bonnet scalar, leading to Einstein-Gauss-Bonnet scalar field gravity.

In the Einstein-Gauss-Bonnet scalar field theory, the gravitational field is
described by the action integral%
\begin{equation}
S_{EGB\phi}=\int d^{4}x\sqrt{-g}\left(  R-\frac{1}{2}g^{\mu\kappa}\nabla_{\mu
}\phi\nabla_{\kappa}\phi-g\left(  \phi\right)  G\right)  , \label{ai.01}%
\end{equation}
where $g\left(  \phi\right)  $ is the coupling function between the scalar
field and the Gauss-Bonnet scalar. Variation with respect to the metric tensor of the action integral
(\ref{ai.01}) leads to the gravitational field equations%
\begin{equation}
R_{\mu\nu}-\frac{1}{2}Rg_{\mu\nu}=T_{\mu\nu}^{G}+T_{\mu\nu}^{\phi},
\label{ai.03}%
\end{equation}
where the energy-momentum tensor $T_{\mu\nu}^{G}~$attributes the
geometrodynamical degrees of freedom provided by the Gauss-Bonnet component
$g\left(  \phi\right)  G$, given by the expression
\begin{align}
T_{\mu\nu}^{G}  &  =-4\left(  \nabla_{\mu}\nabla_{\nu}g\left(  \phi\right)
\right)  R+8\left(  \nabla_{\mu}\nabla_{\rho}g\left(  \phi\right)  \right)
R_{\nu}^{\rho}+8\left(  \nabla_{\nu}\nabla_{\rho}g\left(  \phi\right)
\right)  R_{\mu}^{\rho}\nonumber\\
&  -8\left(  g^{\kappa\rho}\nabla_{\kappa}\nabla_{\rho}g\left(  \phi\right)
\right)  \left(  4R_{\mu\nu}-2Rg_{\mu\nu}\right)  -8\left(  \nabla_{\kappa
}\nabla_{\rho}g\left(  \phi\right)  \right)  \left(  R^{\rho\kappa}g_{\mu\nu
}-R_{\mu~~\nu}^{~\ \rho~~\sigma}\right)  , \label{ai.04}%
\end{align}
and $T_{\mu\nu}^{\phi}$ attributes the scalar field kinetic component, that
is,%
\[
T_{\mu\nu}^{\phi}=\nabla_{\mu}\phi\nabla_{\nu}\phi-g_{\mu\nu}\left(  \frac
{1}{2}g^{\kappa\rho}\nabla_{\kappa}\phi\nabla_{\rho}\phi\right)
\text{\thinspace}.
\]
Furthermore, variation with respect to the scalar field leads to the modified
Klein-Gordon equation
\begin{equation}
-g^{\mu\nu}\nabla_{\mu}\nabla_{\nu}\phi+g_{,\phi}G=0. \label{ai.05}%
\end{equation}
In the following we consider that $g\left(  \phi\right)  =\alpha\phi$ is a
linear function. In this case the energy-momentum tensor $T_{\mu\nu}^{G}$
reads%
\begin{align}
T_{\mu\nu}^{G}  &  =-4\alpha_{0}\left(  \nabla_{\mu}\nabla_{\nu}\phi\right)
R+8\alpha_{0}\left(  \nabla_{\mu}\nabla_{\rho}\phi\right)  R_{\nu}^{\rho
}+8\alpha_{0}\left(  \nabla_{\nu}\nabla_{\rho}\phi\right)  R_{\mu}^{\rho
}\nonumber\\
&  -8\alpha_{0}\left(  g^{\kappa\rho}\left(  \nabla_{\kappa}\nabla_{\rho}%
\phi\right)  \right)  \left(  4R_{\mu\nu}-2Rg_{\mu\nu}\right)  -8\alpha
_{0}\left(  \nabla_{\kappa}\nabla_{\rho}\phi\right)  \left(  R^{\rho\kappa
}g_{\mu\nu}-R_{\mu~~\nu}^{~\ \rho~~\sigma}\right)  ,
\end{align}
and the equation of motion for the scalar field is simplified to
\begin{equation}
-g^{\mu\nu}\nabla_{\mu}\nabla_{\nu}\phi+\alpha_{0}G=0.
\end{equation}

\subsection{Bianchi I spacetime}

In order to investigate the evolution of  anisotropies in
Einstein-Gauss-Bonnet scalar field theory, we consider the Bianchi I spacetime
with the line element \cite{Ryan:1975jw}
\begin{equation}
ds^{2}=-N^{2}\left(  t\right)  dt^{2}+\left(  S_{1}\left(  t\right)  \right)
^{2}dx^{2}+\left(  S_{2}\left(  t\right)  \right)  ^{2}dy^{2}+\left(
S_{3}\left(  t\right)  \right)  ^{2}dz^{2},
\end{equation}
which is the simplest anisotropic model, where each space direction is an
isometry, and the spacetime is invariant under a three-dimensional Abelian
translation group.

Function $N\left(  t\right)  $ is the lapse function and $S_{1}\left(
t\right)  ,~S_{2}\left(  t\right)  $ and $S_{3}\left(  t\right)  $ are the
three scale factors, which define the volume of the three-dimensional
hypersurface $V=S_{1}\left(  t\right)  S_{2}\left(  t\right)  S_{3}\left(
t\right)  $. The three expansions rates (Hubble functions) are defined as 
\begin{equation}
H_{1}=\frac{\dot{S}_{1}}{S},~H_{2}=\frac{\dot{S}_{2}}{S_{2}},~H_{3}=\frac
{\dot{S}_{3}}{S_{3}}. \label{ai.03a}%
\end{equation}

In the case of GR, the vacuum Bianchi I geometry reveals to
the analytic Kasner universe \cite{Kasner:1921zz} an important closed-form solution with many applications in the description of the gravitational field. The Kasner metric depends on three parameters, also known as the Kasner indices, which are constrained by the Kasner algebraic relations \cite{Ryan:1975jw}.  The values of the parameters are defined on the real number line by the intersection of a three-dimensional sphere of radius unity, and a plane in which the sum of those parameters is one. However, the introduction of the cosmological constant makes the Kasner solution unstable.

For our analysis, we elect to work in the Misner variables \cite{Ryan:1975jw} where the line element (\ref{ai.03}) is given by%
\begin{equation}
ds^{2}=-N^{2}\left(  t\right)  dt^{2}+e^{2a}\left(  e^{2\beta_{+}\left(
t\right)  }dx^{2}+e^{-\beta_{+}\left(  t\right)  }\left(  e^{\beta_{-}\left(
t\right)  }dy^{2}+e^{-\beta_{-}\left(  t\right)  }dz^{2}\right)  \right).
\end{equation}
The volume is defined as $V\left(  t\right)  =e^{3a\left(  t\right)  }$, such
that the expansion rate, i.e. the Hubble function is to be defined as usual
$H=\dot{a}$. Functions $\beta_{+}\left(  t\right)  $,~$\beta_{-}\left(
t\right)  $ define the anisotropic parameters $\sigma_{+}=\dot{\beta}_{+}$ and
$\sigma_{-}=\dot{\beta}_{-}$. In the limit these two parameters are zero and the
spacetime takes the form of the the spatially flat FLRW geometry.

The relations between the Hubble function $H~$ and the anisotropic parameters
$\Sigma_{+}$ and $\Sigma_{-}$ with the three expansion rates $H_{1}\,,~H_{2}$
and $H_{3}$ of the Killing directions are as follows
\begin{equation}
H=\frac{1}{3}\left(  H_{1}+H_{2}+H_{3}\right)  , \label{ai.04a}%
\end{equation}
and%
\begin{align}
H_{1}  &  =H\left(  1+\Sigma_{+}\right)  ,~\\
H_{2}  &  =H\left(  1-\frac{1}{2}\left(  \Sigma_{+}-\sqrt{3}\Sigma_{-}\right)
\right)  ,\\
H_{3}  &  =H\left(  1-\frac{1}{2}\left(  \Sigma_{+}+\sqrt{3}\Sigma_{-}\right)
\right)  .
\end{align}

\subsection{Minisuperspace description}

The gravitational field equations which describe the dynamical evolution of
the scale factors $a\left(  t\right)  ,~\beta_{+}\left(  t\right)  $ and
$\beta_{-}\left(  t\right)  $, within the Einstein-Gauss-Bonnet scalar field
theory admit a minisuperspace description. From the line element (\ref{ai.04})
we can calculate the Ricci scalar and the Gauss-Bonnet scalar as
\begin{equation}
R=6\ddot{a}+12\dot{a}^{2}+\frac{3}{2}\left(  \dot{\beta}_{+}\right)
^{2}+\frac{1}{2}\left(  \dot{\beta}_{-}\right)  ^{2},
\end{equation}%
\begin{equation}
\int Ne^{3a}Gdt=\frac{2}{N^{3}}e^{3a}\left(  \dot{a}+\dot{\beta}_{+}\right)
\left(  2\dot{a}-\dot{\beta}_{+}-\dot{\beta}_{-}\right)  \left(  2\dot{a}%
-\dot{\beta}_{-}+\dot{\beta}_{+}\right). \label{ai.06}%
\end{equation}

We replace the latter expressions in (\ref{ai.01}) and after integration by
parts we derive the point-like Lagrangian function
\begin{align}
L\left(  N,a,\dot{a},\beta_{\pm},\dot{\beta}_{\pm}\right)   &  =\frac{e^{3a}%
}{N}\left(  -3\dot{a}^{2}+\frac{3}{4}\dot{\beta}_{+}^{2}+\frac{1}{4}\dot
{\beta}_{-}^{2}-\frac{1}{2}\dot{\phi}^{2}\right)  \nonumber\\
&  ~~~~~+\left(  \frac{e^{3a}}{N^{3}}\alpha_{0}\dot{\phi}\left(  \dot
{a}+\dot{\beta}_{+}\right)  \left(  2\dot{a}-\dot{\beta}_{+}-\dot{\beta}%
_{-}\right)  \left(  2\dot{a}-\dot{\beta}_{-}+\dot{\beta}_{+}\right)  \right)
.\label{ai.07}%
\end{align}
where we have assumed $g\left(  \phi\right)  =\alpha_{0}\phi$ and the kinetic term $-\frac12\dot{\phi}^2$ is present. Hence, the gravitational field equations are%
\begin{align}
0  &  =-3H^{2}+\frac{3}{4}\sigma_{+}^{2}+\frac{1}{4}\sigma_{-}^{2}-\frac{1}%
{2}\dot{\phi}^{2}\nonumber\\
&  +3\alpha_{0}\dot{\phi}\left(  H+\sigma_{+}\right)  \left(  2H-\sigma
_{+}-\sigma_{-}\right)  \left(  2H-\sigma_{-}+\sigma_{+}\right)  ,
\end{align}%
\begin{align}
0  &  =2\left(  1-4a_{0}H\dot{\phi}\right)  \dot{H}+\frac{3}{4}\sigma_{+}%
^{2}+\frac{1}{4}\sigma_{-}^{2}+H^{2}\left(  3-4a_{0}\ddot{\phi}\right)
-8a_{0}H^{3}\dot{\phi}\nonumber\\
&  +\frac{1}{3}\alpha_{0}\ddot{\phi}\left(  3\sigma_{+}^{2}+\sigma_{-}%
^{2}\right)  +\frac{1}{6}\left(  6\alpha_{0}\sigma_{+}\left(  \sigma_{+}%
^{2}-\sigma_{-}^{2}+2\dot{\sigma}_{+}\right)  +4\alpha_{0}\sigma_{-}%
\dot{\sigma}_{-}-3\dot{\phi}\right)  \dot{\phi},
\end{align}%
\begin{align}
0  &  =-\frac{3}{2}\dot{\sigma}_{+}+2\alpha_{0}\dot{\phi}\left(  3\sigma
_{+}\left(  \dot{H}-\dot{\sigma}_{+}\right)  +\sigma_{-}\dot{\sigma}%
_{-}+90H^{2}\sigma_{+}\right) \nonumber\\
&  +\alpha_{0}\ddot{\phi}\left(  \sigma_{-}^{2}-3\sigma_{+}^{2}\right)
+\frac{3}{2}H\left(  2\alpha_{0}\dot{\phi}\left(  2\dot{\sigma}_{+}^{2}%
+\sigma_{-}^{2}\right)  +\sigma_{+}\left(  4\alpha_{0}\ddot{\phi}-3\right)
-6\alpha_{0}\sigma_{+}^{2}\dot{\phi}\right)  ,
\end{align}%
\begin{align}
0  &  =\left(  -1+4\alpha_{0}\sigma_{+}\dot{\phi}\right)  \dot{\sigma}%
_{-}+12\alpha_{0}\sigma_{-}\dot{\phi}+4\alpha_{0}\sigma_{-}\left(  \dot{\phi
}\left(  \dot{H}+\dot{\sigma}_{+}\right)  +\sigma_{+}\ddot{\phi}\right)
\nonumber\\
&  +H\left(  4\alpha_{0}\dot{\phi}\dot{\sigma}_{-}+\sigma_{-}\left(
4\alpha_{0}\left(  \ddot{\phi}+3\sigma_{+}\dot{\phi}\right)  -3\right)
\right)  ,
\end{align}%
\begin{align}
0  &  =\ddot{\phi}-3\alpha_{0}H^{2}\left(  4H^{2}-\left(  3\sigma_{+}%
^{2}+\sigma_{-}^{2}-4\dot{H}\right)  \right)  +3\alpha_{0}\sigma_{+}%
^{2}\left(  \dot{H}-\dot{\sigma}_{+}\right)  +\alpha_{0}\sigma_{-}^{2}\left(
\dot{H}+\dot{\sigma}_{+}\right) \nonumber\\
&  +2\alpha_{0}\sigma_{+}\sigma_{-}\dot{\sigma}_{-}+3H\left(  \dot{\phi
}+\alpha_{0}\sigma_{+}\left(  \sigma_{-}^{2}-\sigma_{+}^{2}+2\dot{\Sigma}%
_{+}\right)  +\frac{2}{3}\alpha_{0}\sigma_{-}\dot{\sigma}_{-}\right)  .
\end{align}
where without loss of generality we have selected $N\left(  t\right)  =1$.

\section{Locally Rotational Spacetime}

\label{sec3}

Assume that the Bianchi I spacetime admits a fourth isometry, which is the
rotation symmetry between the $y$ and $z$ directions. That means that the
scale factors $S_{2}\left(  t\right)  =S_{3}\left(  t\right)  $, and in the
Misner variables, $\sigma_{-}=0$. We introduce the dimensionless variables%
\begin{equation}
x=\phi\sqrt{1+H^{2}},~\Sigma_{+}=\frac{\sigma_{+}}{\sqrt{1+H^{2}}},~\eta
=\frac{H}{\sqrt{1+H^{2}}},~d\tau=\sqrt{1+H^{2}}dt,\label{ai.08}%
\end{equation}
where now the field equations are expressed by a system of first-order
differential equations%
\begin{align}
\frac{dx}{d\tau} &  =f_{1}\left(  x,\Sigma_{+},\eta\right)  ,\label{ai.09}\\
\frac{d\Sigma_{+}}{d\tau} &  =f_{2}\left(  x,\Sigma_{+},\eta\right)
,\label{ai.10}\\
\frac{d\eta}{d\tau} &  =f_{3}\left(  x,\Sigma_{+},\eta\right)  ,\label{ai.11}%
\end{align}
with the algebraic constraint%
\begin{equation}
3\eta^{2}+\frac{1}{2}\left(  \eta^{2}-1\right)  ^{2}x^{2}-\frac{3}{4}%
\Sigma_{+}^{2}-3\alpha_{0}x\left(  \Sigma_{+}-2\eta\right)  ^{2}\left(
\Sigma_{+}+\eta\right)  =0.\label{ai.12}%
\end{equation}

We proceed with the analysis of the stationary points of the dynamical system
(\ref{ai.09})-(\ref{ai.12}). Such analyses reveal important information about
the asymptotic solutions of the gravitational model and their stability properties. Indeed, the stationary points $P=\left(  x\left(  P\right)  ,\Sigma_{+}\left(
P\right)  ,\eta\left(  P\right)  \right)  $ of the gravitational field
equations\ are
\begin{align*}
P_{1}^{\pm} &  =\left(  \pm\frac{\sqrt{9-6\sqrt{30}\alpha_{0}}}{10\alpha_{0}%
},0,\pm\frac{3}{\sqrt{9-6\sqrt{30}\alpha_{0}}}\right)  ,\\
P_{2}^{\pm} &  =\left(  \pm\frac{\sqrt{9+6\sqrt{30}\alpha_{0}}}{10\alpha_{0}%
},0,\pm\frac{3}{\sqrt{9+6\sqrt{30}\alpha_{0}}}\right)  ,\\
P_{3}^{\pm} &  =\left(  \pm\frac{\sqrt{4-9\sqrt{2}\alpha_{0}}}{6\alpha_{0}%
},\pm\frac{1}{\sqrt{4-9\sqrt{2}\alpha_{0}}},\pm\sqrt{\frac{8+18\sqrt{2}%
\alpha_{0}}{8-81\alpha_{0}^{2}}}\right)  \\
P_{4}^{\pm} &  =\left(  \pm\frac{\sqrt{4+9\sqrt{2}\alpha_{0}}}{6\alpha_{0}%
},\pm\frac{1}{\sqrt{4+9\sqrt{2}\alpha_{0}}},\pm\sqrt{\frac{8-18\sqrt{2}%
\alpha_{0}}{8-81\alpha_{0}^{2}}}\right)  \\
P_{5} &  =\left(  0,0,0\right)
\end{align*}
Points $P_{1}^{\pm}$ and $P_{2}^{\pm}~$ are real and physically acceptable when
$\alpha_{0}<\frac{1}{2}\sqrt{\frac{3}{10}}$, and $\alpha_{0}>-\frac{1}{2}%
\sqrt{\frac{3}{10}}$ respectively. On the other hand, points $P_{3}^{\pm}$
and $P_{4}^{\pm}$ are physically acceptable when $\left\{  \alpha_{0}<\frac
{2\sqrt{2}}{9},\alpha_{0}\neq-\frac{2\sqrt{2}}{9}\right\}  $ and $\left\{
\alpha_{0}>-\frac{2\sqrt{2}}{9},~\alpha_{0}\neq\frac{2\sqrt{2}}{9}\right\}  $. 

In order to understand the physical properties of the asymptotic solutions at
the stationary points, we calculate the deceleration parameters$~q=-1-\frac
{\dot{H}}{H^{2}}$ at the points. We find that the parameters for the points
$P_{1}^{\pm},~P_{2}^{\pm},~P_{3}^{\pm}$ and $P_{4}^{\pm}$  have value $-1$.
This means that at the asymptotic solutions, the Hubble function $H\left(
t\right)$ is constant, that is $H\left(  t\right)  =H_{0}$. This is an
expected result, because by definition $H=\frac{\eta}{\sqrt{1-\eta^{2}}}$,
hence for $\left\vert \eta\right\vert$ at a constant different from one, the later
expression reveals that Hubble function is a constant. For point $P_{5}$ the
deceleration parameter is not defined. 

Therefore, the stationary points $P_{1}^{\pm}\,\ $ and $P_{2}^{\pm}$ describe
de Sitter FLRW geometries, while points $P_{3}^{\pm}$ and $P_{4}^{\pm}$
describe anisotropic exponential expansion. It is notable that point $P_{5}$ describes the Minkowski solution. We remark that we have not introduced a cosmological constant in the gravitational theory. The scalar field and the Gauss-Bonnet term play the role of inflation, which drives the dynamics to describe
inflation. We remark that scaling solutions are not supported by the
gravitational model.

The stability of the stationary points are investigated numerically. We find
that $P_{1}^{+}$ is an attractor for $\alpha_{0}<0$. Also,  $P_{1}^{-}$, $P_{2}^{+}$ and $P_{2}^{-}$ describe unstable solutions. Moreover, $P_{3}^{+}$ is an
attractor for $\alpha_{0}<0$ while $P_{4}^{-}$ is an attractor for $-\frac
{2\sqrt{2}}{9}<\alpha_{0}<0$, and $P_{3}^{-}$,~$P_{4}^{+}$ describe unstable
solutions. Finally, point $P_{5}$ corresponds to an unstable solution.
Specifically $P_{5}$ describes the transition of the trajectories within the
phase-space from the region of $H>0$ to the region where $H<0$, and vice
versa. After considering the LRS case, we can now investigate anisotropies more generally.

\section{Compactification }

\label{sec4}

We proceed with the analysis of the evolution of the anisotropies in the more
general case where $\sigma_{-}$ is now a dynamical variable. For this model, we
introduce the additional dimensionless parameter
\[
\Sigma_{-}=\frac{\sigma_{-}}{\sqrt{1+H^{2}}}.
\]
Due to the nonlinearity of the field equations, the resulting dynamical system
is studied numerically. We employ the constraint equation and we reduce the
dimension of the dynamical system by one. In particular, we explore the
numerical solutions in the two branches $x_{+}$ and $x_{-}$ as they come 
from the constraint equation. 

We performed numerical simulations for various sets of initial conditions
for the dynamical variables $\left\{  \Sigma_{1},\Sigma_{2},\eta\right\}  $,
and for the free parameter $\alpha_{0}$. In the following, we present and
discuss the behavior of dynamical variables for solution trajectories where
the dynamical variables remain in the finite regime.  

\begin{figure}[ptbh]
\centering\includegraphics[width=1\textwidth]{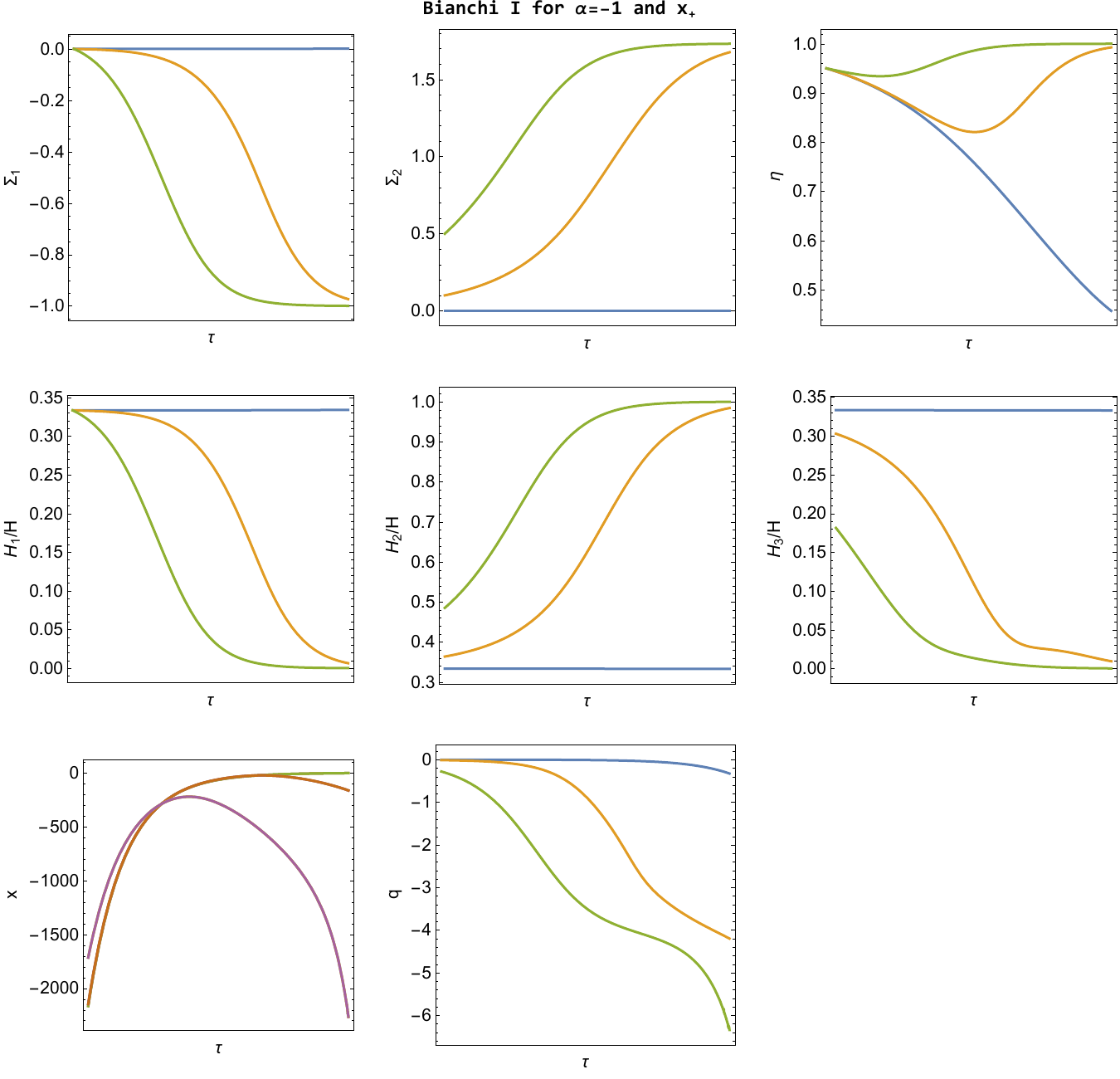}\caption{Qualitative
evolution in the branch $x_{+}$ for $\alpha=-1$ for the anisotropic parameters
\thinspace$\Sigma_{1},~\Sigma_{2}$, of the deceleration parameter $q$, of the
scalar field $x$, of the Hubble parameters $H_{1},~H_{2}$ and $H_{3}$ and of
parameter $\eta$ for the initial conditions $\Sigma_{1}^{0}=0$,~$\Sigma
_{2}^{0}=\left(  0,0.1,0.5\right)  $ and $\eta=0.95$. }%
\label{num1}%
\end{figure}

\begin{figure}[ptbh]
\centering\includegraphics[width=1\textwidth]{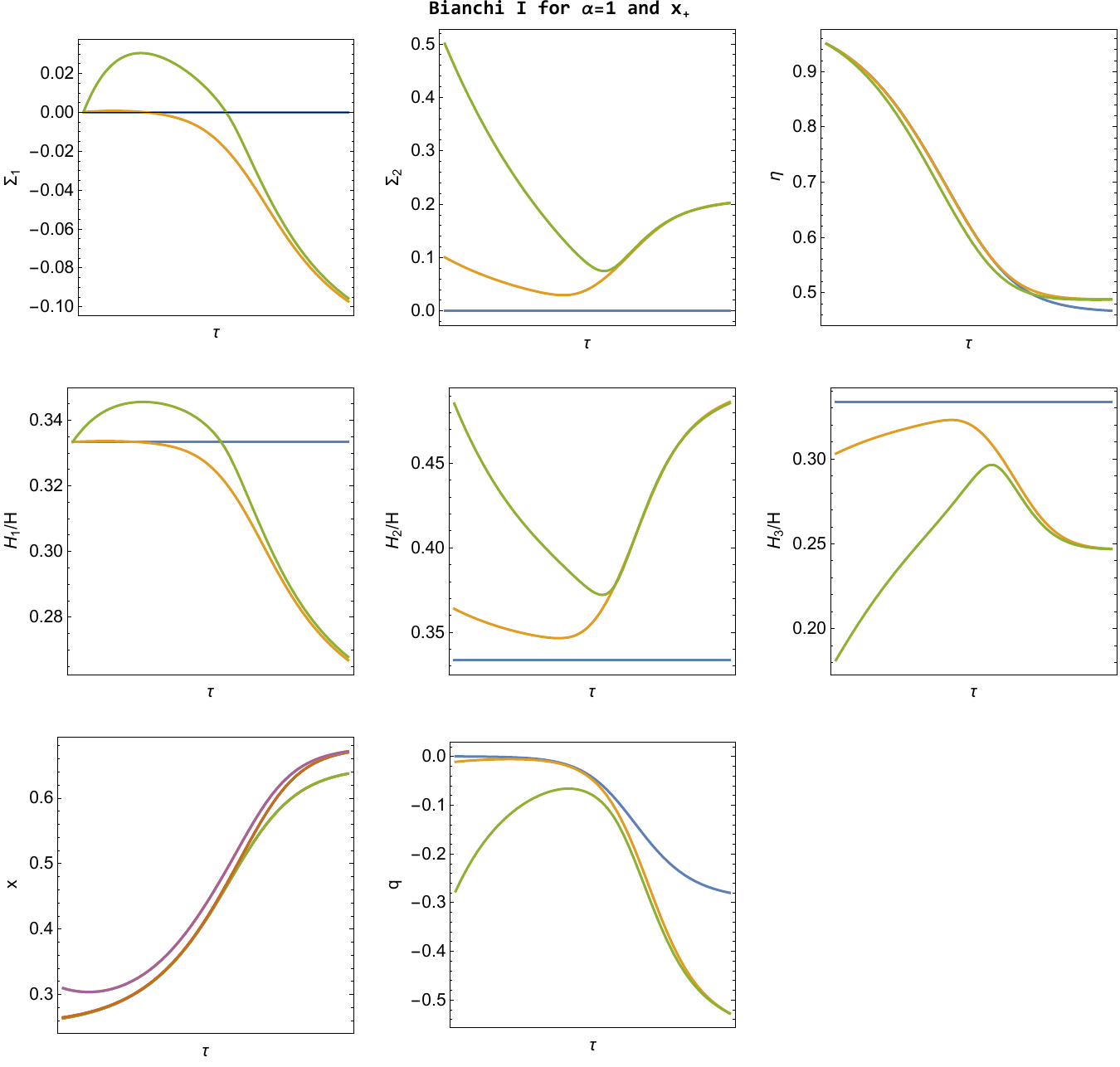}\caption{Qualitative
evolution in the branch $x_{+}$ for $\alpha=1$ for the anisotropic parameters
\thinspace$\Sigma_{1},~\Sigma_{2}$, of the deceleration parameter $q$, of the
scalar field $x$, of the Hubble parameters $H_{1},~H_{2}$ and $H_{3}$ and of
parameter $\eta$ for the initial conditions $\Sigma_{1}^{0}=0$,~$\Sigma
_{2}^{0}=\left(  0,0.1,0.5\right)  $ and $\eta=0.95$. }%
\label{num2}%
\end{figure}

\begin{figure}[ptbh]
\centering\includegraphics[width=1\textwidth]{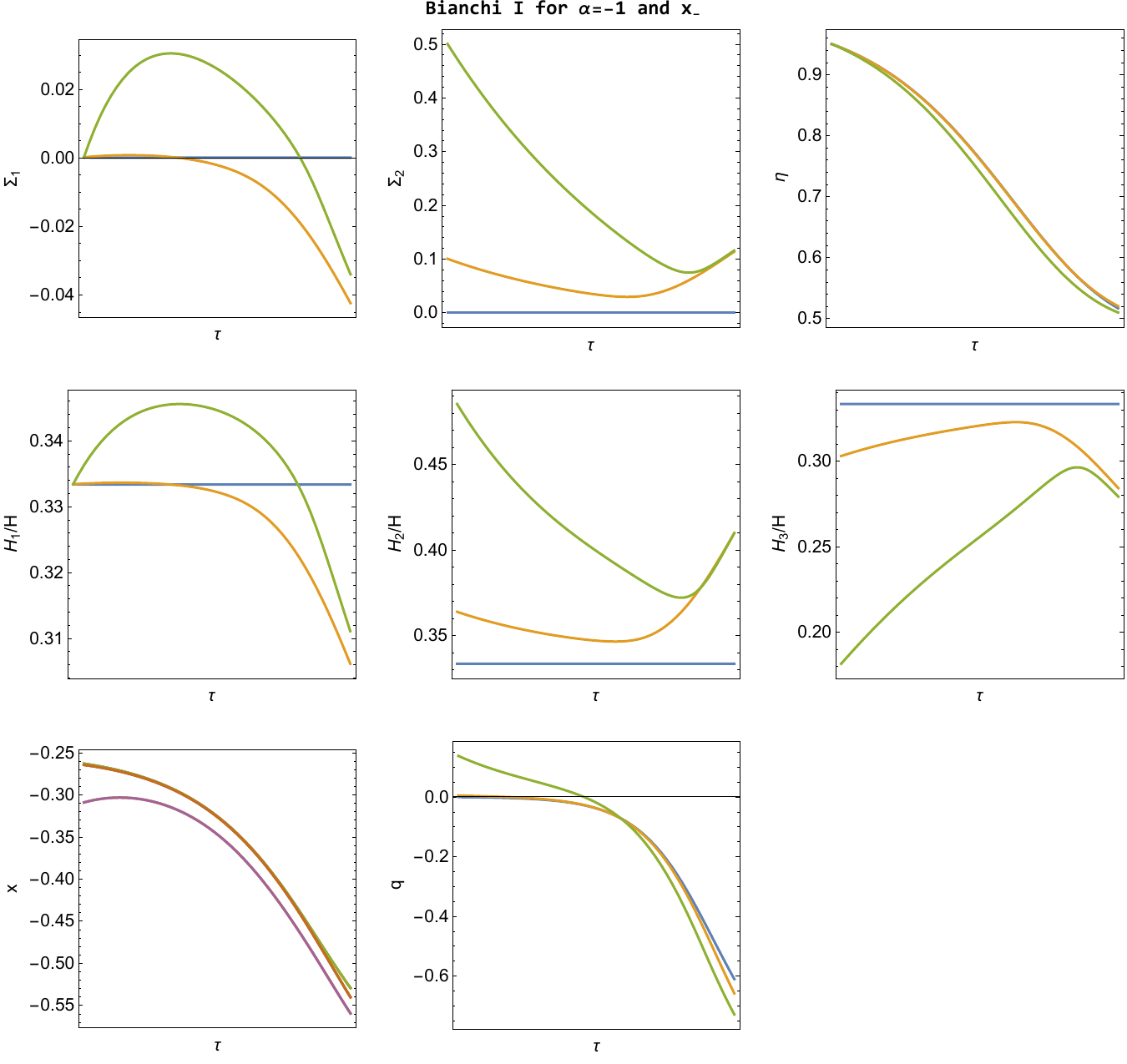}\caption{Qualitative
evolution in the branch $x_{-}$ for $\alpha=-1$ for the anisotropic parameters
\thinspace$\Sigma_{1},~\Sigma_{2}$, of the deceleration parameter $q$, of the
scalar field $x$, of the Hubble parameters $H_{1},~H_{2}$ and $H_{3}$ and of
parameter $\eta$ for the initial conditions $\Sigma_{1}^{0}=0$,~$\Sigma
_{2}^{0}=\left(  0,0.1,0.5\right)  $ and $\eta=0.95$. }%
\label{num3}%
\end{figure}

\begin{figure}[ptbh]
\centering\includegraphics[width=1\textwidth]{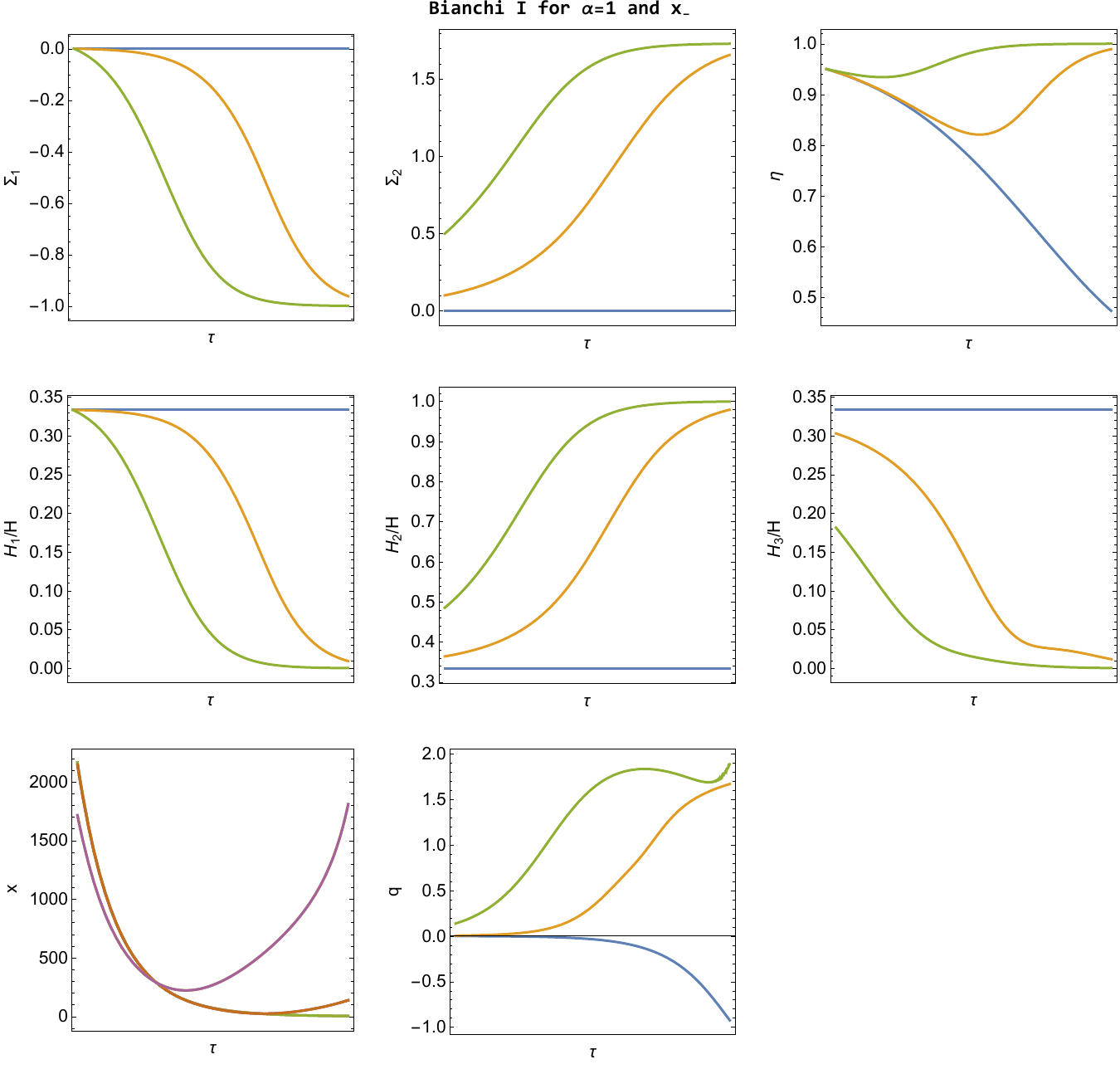}\caption{Qualitative
evolution in the branch $x_{-}$ for $\alpha=1$ for the anisotropic parameters
\thinspace$\Sigma_{1},~\Sigma_{2}$, of the deceleration parameter $q$, of the
scalar field $x$, of the Hubble parameters $H_{1},~H_{2}$ and $H_{3}$ and of
parameter $\eta$ for the initial conditions $\Sigma_{1}^{0}=0$,~$\Sigma
_{2}^{0}=\left(  0,0.1,0.5\right)  $ and $\eta=0.95$. }%
\label{num4}%
\end{figure}

In Figs \ref{num1}, \ref{num2}, \ref{num3} and \ref{num4} we present the
trajectories which lead to physically viable solutions within the finite
regime of the dynamical variables. 

Fig. \ref{num1} depicts the qualitative evolution for the branch
$x_{+}$ with $a=-1$. The blue line is for initial conditions very close to
the isotropic geometry, while the other two lines describe initial conditions
with higher anisotropic parameters. We remark that the de Sitter solution is
an attractor for the dynamical system. The other two solutions given by the green and orange lines describe scaling solutions, that is Kanser-like universes, where the two scale factors $S_{2}$ and $S_{3}$ are constant. However, from the behavior of the deceleration parameter and of the scalar fields, for these two trajectories, we conclude that the spacetime is compactified to a $2+2$ geometry. That is, the $x$- and $z$- directions are compactified.  We remark that similar behaviors are achieved for $a<0$. Also a similar behaviour is achieved  for the branch $x_{-}$ for $a=+1$ as presented in Fig. \ref{num3}

In Fig. \ref{num2} we present the trajectories in the branch $x_{+}$ for
$a=+1$. We observe that the FLRW geometry is an attractor. Moreover, the
rest of the trajectories describe expanding solutions where $\eta$ is a
constant. A similar behavior is recovered in Fig. \ref{num3} which describes
solutions in the branch $x_{-}$ and $a=-1$.

\section{Conclusions}

\label{sec5}

In this work we studied the evolution of anisotropies in Bianchi I cosmologies in the four-dimensional Einstein-Gauss-Bonnet scalar field theory with a canonical kinetic term and a linear coupling $g(\phi)=\alpha_{0}\phi$. By introducing a convenient set of dimensionless variables, we reformulated the field equations as an autonomous dynamical system and examined the asymptotic behavior using a combination of analytical tools (for the LRS case) and numerical integration (for the fully anisotropic case). 

For the LRS Bianchi I spacetime we found that the physically relevant stationary points correspond to exponential expansion with $q=-1$. These include isotropic de
Sitter-like solutions as well as anisotropic exponential solutions, while Minkowski spacetime corresponds to an unstable stationary point that can rather be viewed as a passage between expansion and contraction, not as a stable late-time attractor. Notably, these accelerating solutions arise without introducing an explicit cosmological constant. Hence we can conclude that the scalar field and the Gauss-Bonnet term together provide an effective vacuum energy that drives the expansion, effectively playing the role of a cosmological constant. In addition, the dynamical system does not support scaling solutions, so the asymptotics are not of power-law type but instead are selected from the de Sitter-like family (isotropic or anisotropic) depending on the sign and range of $\alpha_{0}$, as supported by numerical investigations we performed.

When the general (full) anisotropic sector is included, our numerical analysis shows that the theory admits even richer asymptotic behavior. Besides trajectories attracted to the isotropic de Sitter solution, we identified Kasner-like regimes associated with compactification, including solutions that realize a $2+2$ effective splitting of spacetime. Moreover, we found a new family of solutions exhibiting a two-dimensional splitting of the spatial geometry, parallel to compactification phenomena known from higher-dimensional pure Einstein-Gauss-Bonnet models. These results indicate that the scalar kinetic term can qualitatively change the anisotropic dynamics. The geometric splitting
persists but it can now become dynamically preferred and act as an attractor for broad sets of initial conditions. Future work includes extending this analysis to more general coupling functions and to the inclusion of a scalar potential.

\textbf{Data Availability Statements:} Data sharing is not applicable to this
article as no datasets were generated or analyzed during the current study.
\newline\newline\textbf{Code Availability Statements:} Code sharing is
available after request. \newline

\section*{Acknowledgments}
AG was supported by Proyecto Fondecyt Regular Number 1240247. GL \& AP were
supported by Proyecto Fondecyt Regular Number 1240514. CH thanks the National Institute for Theoretical and Computational Sciences (NiTheCS) for a writing retreat grant. AP thanks CH for the hospitality provided when this work was carried out.

\bibliography{biblio}
\end{document}